\documentclass[%
 reprint,
 amsmath,amssymb,
 aps,
]{revtex4-2}

\usepackage{amsmath,amssymb,amsfonts,braket}
\usepackage{graphicx}
\usepackage{dcolumn}
\usepackage{bm}
\usepackage[hidelinks,colorlinks=true,linkcolor=blue,urlcolor=blue, citecolor=blue]{hyperref}

\usepackage[dvipsnames]{xcolor}

\newcommand{\upg}{\uparrow_\text{g}}
\newcommand{\downg}{\downarrow_\text{g}}
\newcommand{\upe}{\uparrow_\text{e}}
\newcommand{\downe}{\downarrow_\text{e}}

\newcommand{\boldOpLe}[1]{\mathbf{{#1}}}
\newcommand{\boldOpSy}[1]{\boldsymbol{{#1}}}

\begin{document}


\title{Cavity-assisted single-shot \textit{T}-center spin readout}

\author{Yu-En Wong}
 \affiliation{Department of Electrical and Computer Engineering, Rice University, Houston, TX 77005, USA}
\affiliation{%
 Applied Physics Graduate Program, Smalley-Curl Institute, Rice University, Houston, TX 77005, USA
}%

\author{Songtao Chen}
    \email{songtao.chen@rice.edu}
    \affiliation{Department of Electrical and Computer Engineering, Rice University, Houston, TX 77005, USA}
    \affiliation{Smalley-Curl Institute, Rice University, Houston, TX 77005, USA}

\date{\today}

\begin{abstract}
High-fidelity spin readout is a crucial component for quantum information processing with optically interfaced solid-state spins. Here, we propose and investigate two theoretical protocols for fast single-shot readout of cavity-coupled single \textit{T} center electronic spins. For fluorescence-based readout, we selectively couple one of the \textit{T} center spin-conserving transitions to a single-mode photonic cavity, exploiting the enhancement of the fluorescence emission and cyclicity. 
For reflection-based readout, we leverage the spin-dependent cavity reflection contrast to generate the qubit readout signal. 
We show that the cavity reflection approach enables high-fidelity spin readout even when the \textit{T} center only has a modest cyclicity. With realistic system parameters, such as cavity quality factor $Q = 2\times10^5$ and \textit{T}-center optical linewidth $\Gamma/2\pi = 100$ MHz, we calculate a single-shot readout fidelity exceeding 99\% within 10 
$\mu$s for both spin readout protocols. 

\end{abstract}

\maketitle


\section{\label{sec:level1_intro}Introduction\protect}

Optically interfaced solid-state spins have been widely used for a variety of quantum technologies \cite{awschalom2018quantum}, including quantum interconnect and networking. These spin systems can store quantum information locally in individually addressable particles and send out spin-entangled photons for networking applications. Using mature solid-state spin platforms such as quantum dots and defect centers in diamond (e.g., NV, SiV), efficient spin-photon entanglement \cite{togan2010quantum, gao2012observation, nguyen2019integrated} and remote spin entanglement \cite{bernien2013heralded,pfaff2014unconditional} have been demonstrated. Quantum frequency conversion is commonly used in these platforms to convert the wavelength to telecom-band for long-range applications \cite{de2012quantum,knaut2024entanglement}. Single erbium ions in oxide materials provide spin-photon interfaces at telecom-C band \cite{dibos2018atomic}, with spin-photon entanglement demonstrated recently \cite{uysal2025spin}.

Radiation damage centers are another family of defects that offer telecom-band optical emission and scalable photonic and electronic device integration in the mature silicon material platform \cite{jones1973temperature, chartrand2018highly}. Single \textit{T} centers are particularly promising optically active spins in silicon (OASIS) for building quantum network nodes and repeater devices due to their telecom-O band optical transition and the long-lived doublet spin manifold in the ground state \cite{bergeron2020silicon}. Several recent works have demonstrated single \textit{T}-center isolation and silicon photonic device integration \cite{higginbottom2022optical,johnston2024cavity,lee2023high,islam2023cavity,komza2025multiplexed}, as well as coherent coupling to the nearby nuclear spins \cite{song2026entanglement}.

An important requirement for quantum information processing using solid-state spins is the high-fidelity single-shot spin readout. Based on the cavity-enhanced fluorescence enabled by the cavity quantum electrodynamics (cQED), single-shot spin measurements have been achieved in a variety of solid-state spin platforms, including quantum dots \cite{vamivakas2010observation, delteil2014observation}, NV \cite{robledo2011high} and SiV \cite{sukachev2017silicon} centers in diamond, as well as rare-earth ions \cite{raha2020optical,kindem2020control}. High-fidelity spin readout based on resonant fluorescence relies on the highly cyclic optical transitions to scatter sufficient photons for detection before spin flips. Cycling transition can be enabled by intrinsic atomic selection rules \cite{delteil2014observation}, tuning transition dipole alignment with an external magnetic field \cite{sukachev2017silicon, rosenthal2024single}, or by tailoring the electromagnetic density of states via a cavity \cite{raha2020optical, kindem2020control}. Leveraging cQED in the intermediate- and high-cooperativity regime, spin-modulated cavity reflection or transmission has been used to perform high-fidelity single-shot spin readout in SiV \cite{evans2018photon, nguyen2019integrated, bhaskar2020experimental}. Moreover, spin-to-charge conversion is also utilized to enable single-shot spin readout \cite{shields2015efficient, anderson2022five}. For \textit{T} centers in silicon, high-fidelity single-shot readout has been demonstrated for the $^1$H nuclear spin \cite{afzal2024distributed}, however, the same readout task for single \textit{T}-center electronic spins remains elusive.

In this work, we propose two cQED protocols to perform single-shot readout of the \textit{T}-center electronic spin. We first construct our theoretical framework based on a four-level system with two ground spin states ($S = 1/2$) forming the qubit, and two Zeeman-split optically excited states, coupled to a low-loss, small mode-volume, and single-mode optical cavity. We then analyze the spin readout performance using two cQED protocols based on resonant fluorescence and cavity reflection. 
Under realistic experimental conditions and system parameters, we calculate the spin readout fidelity by thresholding the probability distribution of spin-dependent photon detections.
We show that a 99.96\% fidelity can be realized in fluorescence-based readout, although it requires high cyclicity, short laser excitation pulses, and high-efficiency detection of fast resonant fluorescence. By using spin-modulated cavity reflection, even with a modest cyclicity, we show that a 99.9935\% readout fidelity can be achieved.

The paper is organized as follows. Section~\ref{sec:level1_Methods} describes the theoretical framework for the protocols; Sections ~\ref{sec:level2_FluorescentReadout} and~\ref{sec:level2_RelectionReadout} analyze the performance of the fluorescence-based and reflection-based spin readout, respectively. In Sec.~\ref{sec:level3_SpectralDiffusion}, we will further investigate the impact of spectral diffusion on the readout fidelity.

\section{\label{sec:level1_Methods}Open quantum system: Lindbladian formalism}

We integrate single \textit{T} centers with a low-loss, small mode-volume one-dimensional (1D) photonic crystal nanobeam cavity fabricated on a silicon-on-insulator wafer. The on-chip light coupling is enabled via an apodized subwavelength grating coupler (GC) through an angle-polished fiber \cite{johnston2024cavity}, with a one-way coupling efficiency $\eta_\text{GC} \sim 50\%$.

\begin{figure}[!h]
    \centering
    \includegraphics[scale=1]{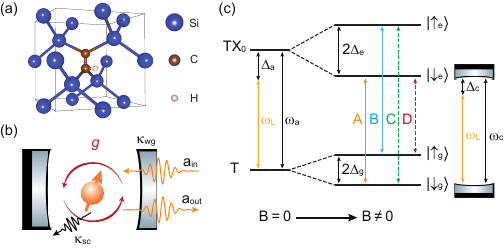}
    \caption{Proposed cQED scheme for the \textit{T}-center electronic spin readout. 
    (a) Schematic of the atomic structure of a single \textit{T} center in the silicon lattice.
    (b) Illustration of a single \textit{T} center coupled to a one-sided cavity, where the input and output laser fields ($\mathbf{a}_\text{in}$, $\mathbf{a}_\text{out}$) couple to the cavity mode at a rate of $\kappa_\text{wg}$. The cavity scattering loss rate $\kappa_\text{sc}$ is also included. The coupling rate between the single \textit{T} center and the cavity mode is given by $g$ at zero field. 
    (c) A simplified energy diagram is shown on the right. Four transitions emerge after an external magnetic field $B$ is applied.
    The Zeeman splittings for ground and excited states are $2\Delta_e$ and $2\Delta_g$, respectively.
    }
    \label{fig:cavity}
\end{figure}

We consider a cavity-coupled single \textit{T} center ($\lambda = 1325.69$ nm) under a static magnetic field, which lifts the spin degeneracy in both the ground and excited states (Fig.~\ref{fig:cavity}), revealing four distinct optical transitions. Transitions \textit{A} and \textit{B} conserve the spin, while \textit{C} and \textit{D} flip the spin. The system Hamiltonian is based on the extended Jaynes-Cummings model, which considers four levels interacting with the single-mode cavity field. The single sided cavity output field [Fig.~\ref{fig:cavity}(b)] is described by the input-output relation (Sec.~\ref{sec:level2_RelectionReadout}).

We write the system Hamiltonian in a reference frame with respect to the frequency of incident field $\omega_\text{L}$, given by $\boldOpLe{H}/\hbar = \boldOpLe{H}_0 + \boldOpLe{H}_\text{int} + \boldOpLe{H}_\textbf{d}$: \vspace{-12pt}

\begin{align}  
    \boldOpLe{H}_0 = &\Delta_c \boldOpLe{a}^{\dag} \boldOpLe{a} + \Delta_g(\boldOpSy{\sigma}_{11}-\boldOpSy{\sigma}_{00}) \notag \\
    & +(\Delta_a-\Delta_e)\boldOpSy{\sigma}_{22}+(\Delta_a+\Delta_e)\boldOpSy{\sigma}_{33},
    \\[2ex]
    \boldOpLe{H}_\text{int} = & g_{\parallel}(\boldOpSy{\sigma}_{31}+\boldOpSy{\sigma}_{20})\boldOpLe{a}^{\dag}
    +g_{\perp}e^{i\phi}(\boldOpSy{\sigma}_{30}+\boldOpSy{\sigma}_{21})\boldOpLe{a}^{\dag} + \text{H.c.},
    \\[2ex]
    \boldOpLe{H}_\text{d} = &\sqrt{\kappa\eta_\text{cav}\epsilon}(\boldOpLe{a}^{\dag}+\boldOpLe{a}),
\end{align}

\noindent where $\ket{0}$, $\ket{1}$, $\ket{2}$, and $\ket{3}$ represent $\ket{\downg}$, $\ket{\upg}$, $\ket{\downe}$, and $\ket{\upe}$ levels [Fig.~\ref{fig:cavity}(c)], respectively;  $\boldOpSy{\sigma}_{ij} = \ket{j}\bra{i}$ with $i,j = 0, 1, 2,3$; $\boldOpLe{a^{\dagger}} (\boldOpLe{a})$ is the photonic creation (annihilation) operators; $\Delta_a = \omega_a-\omega_\text{L}$ and $\Delta_c = \omega_c-\omega_\text{L}$ are the detunings between the bare atomic~($\omega_a$), cavity ($\omega_c$) and laser ($\omega_\text{L}$) frequencies; $2\Delta_g$ and $2\Delta_e$ are the ground and excited state Zeeman splitting; $g_{\parallel}$ and $g_{\perp}$ are the atom-cavity coupling strengths for spin-conserving and spin-non-conserving transitions, with their ratio defined as $r_g =  g_{\parallel}/g_{\perp}$. The phase difference $\phi$ between two couplings originates from the alignment between the external magnetic field and the \textit{T} center spin quantization axis. Our calculations suggest a negligible influence of this phase on the readout fidelity and we keep $\phi=\pi/2$ in the fidelity analysis. 
$2g = 2\sqrt{g_\parallel^2+g_\perp^2}$ is the single-photon Rabi frequency for the bare \textit{T} center; $\epsilon = P_\text{in}/(\hbar\omega_\text{L})$ is the input excitation photon flux, where $P_\text{in}$ is the input laser power right before entering the cavity; $\kappa$ is the total cavity loss rate and $\eta_\text{cav} = \kappa_\text{wg}/\kappa$ is the efficiency for cavity decay into the waveguide ($\kappa_\text{wg}$). Due to the small hyperfine interaction compared to the intrinsic and the cavity-enhanced optical linewidths, we ignore $^1$H nuclear spin levels in the model.

The time evolution of the system dynamics is described by Lindblad master equation
\vspace{-5.5pt}
\begin{equation}
    \label{eq:lindbladian}
    \dot\rho=\frac{-i}{\hbar}[\boldOpLe{H},\rho] +\sum_i{\boldOpLe{c}_i\rho \boldOpLe{c}_i^\dagger - \frac{1}{2}\{\boldOpLe{c}_i^\dagger \boldOpLe{c}_i,\rho \}},
    \vspace{-5.5pt}
\end{equation}
where $\boldOpLe{c}_i$ are the collapse operators describing the interaction with the environment (Table \ref{tab:jumpOperators}). 

\begin{table}[!h]
    \caption{\label{tab:jumpOperators}%
    Collapse operators considered in the Lindbladian. We use balanced transition rates for $\boldOpSy{\sigma}_{20(31)}$ and $\boldOpSy{\sigma}_{21(30)}$ \cite{higginbottom2022optical} in the calculation. $\Gamma_\text{d}$ is the pure dephasing rate.
    }
    \begin{ruledtabular}
        \begin{tabular}{lcdr}
            \textrm{Loss channel}&
            \textrm{Operator} \\
            \colrule
            Cavity decay &  $\sqrt{\kappa}\boldOpLe{a}$ \\
            Excited state dephasing &  $\sqrt{\Gamma_\text{d}/2}\boldOpSy{\sigma}_{z1}$;$\sqrt{\Gamma_\text{d}/2}\boldOpSy{\sigma}_{z2}$
            \footnote{$\boldOpSy{\sigma}_{z1} = \boldOpSy{\sigma}_{22}-\boldOpSy{\sigma}_{00}; \boldOpSy{\sigma}_{z2} = \boldOpSy{\sigma}_{33}-\boldOpSy{\sigma}_{11}$} \\
            Spin-conserving SPE\footnote{Spontaneous emission} &  $\sqrt{\Gamma_0/2}\boldOpSy{\sigma}_{31}$;$\sqrt{\Gamma_0/2}\boldOpSy{\sigma}_{20}$ \\
            Spin-non-conserving SPE  &  $\sqrt{\Gamma_0/2}\boldOpSy{\sigma}_{30}$; $\sqrt{\Gamma_0/2}\boldOpSy{\sigma}_{21}$\\
        \end{tabular}
    \end{ruledtabular}
\end{table}

\noindent To investigate the dynamics of the open quantum system, we solve the master equation numerically with the system Hamiltonian shown above with QuTip \cite{johansson2012qutip}. The obtained atomic and cavity excitations are used to calculate the detected photons and infer the spin readout fidelity. The optical linewidth $\Gamma=\Gamma_0+2\Gamma_\text{d}$ consists of a Fourier-limited linewidth $\Gamma_0 = 2\pi\times169.3$ kHz  \cite{bergeron2020silicon} and pure dephasing $\Gamma_\text{d}$. Extra linewidth broadening can result from spectral diffusion and affect spin readout performance (Sec.~\ref{sec:level3_SpectralDiffusion}). 
Based on this theoretical framework, we investigate two cQED spin readout protocols via resonant fluorescence (Sec.~\ref{sec:level2_FluorescentReadout}) and spin-dependent cavity reflection (Sec.~\ref{sec:level2_RelectionReadout}).

\section{\label{sec:level2_FluorescentReadout}Spin readout via resonant fluorescence}

The branching ratio for bulk \textit{T} centers is relatively balanced \cite{higginbottom2022optical}, leading to a low cyclicity and fast electronic spin polarization under optical excitations. 
We integrate single \textit{T} centers with a low-loss, small mode-volume optical cavity, which can selectively enhance the spin-conserving optical decay pathways \cite{raha2020optical}, enabling cyclic transitions to collect sufficient fluorescence photons before spin flips. 

\begin{figure}[h]
    \centering
    \includegraphics[scale=1]{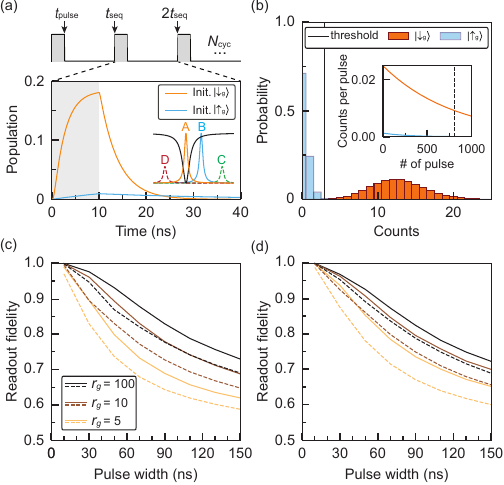}
    \caption{Spin-dependent fluorescence readout.  
    (a) Excited state population dynamics with different initialized states.  The gray shaded region represents the square laser pulse with the readout sequence shown in the top panel. The inset shows the cavity (black line) alignment with the transition \textit{A}.
    (b) Poissonian distribution $P(N_\text{ph},k)$ of detected fluorescence photon counts under different spin initializations. The inset shows the average photon counts per pulse decay due to the optical pumping. The number of excitation pulses $N_\text{cyc}$ for readout is chosen to be $1/e$ decay point (black dashed line).
    Simulation parameters for panels (a), (b): $Q=1\times10^5$, $\Gamma/2\pi=1$ GHz, $t_\text{pulse}$ $=10$ ns, $P_\text{in} = 100$~pW, and $r_g=10$. Both the cavity and the laser are tuned resonant with the transition \textit{A}.
    (c), (d) $r_g$ dependence of readout fidelity with (c) cavity $Q = 1\times10^5$  and (d) $Q=2\times10^5$ . In both panels, solid and dashed lines represent $\Gamma/2\pi=0.1$ GHz and $\Gamma/2\pi=1$ GHz, respectively. Other system parameters can be found in Table \ref{tab:commonParametersFluorescentReadout}.
    }
    \label{fig:Fig2_fluorescenceReadout} 
\end{figure}

To perform the \textit{T} center spin readout via resonant fluorescence, we align the cavity and the resonant laser excitation with one of the spin-conserving transitions, e.g., transition \textit{A} [Fig.~\ref{fig:Fig2_fluorescenceReadout}(a), inset]. A short laser excitation (pulse width of $t_\text{pulse}$) resonantly excites the \textit{T} center and the fluorescence is subsequently collected for a duration of $t_\text{wait} = 7\tau_\text{cav}^\text{off}$, where $\tau_\text{cav}^\text{off}$ is the fluorescence lifetime for the other spin-conserving transition that is off-resonant with the cavity. This practice is chosen to ensure the excited-state population fully decays after each excitation pulse regardless of the initialized spin state, which enables us to extract the cyclicity accurately. We repeat the excite-collect sequence (length of $t_\text{seq} = t_\text{pulse} + t_\text{wait}$) $N_\text{cyc}$ (defined below) times to collect fluorescence photons for spin readout. Depending on the \textit{T}-center spin state, different intensity of fluorescence will be obtained [Fig.~\ref{fig:Fig2_fluorescenceReadout}(a)]. 
We gate our single photon detector in the time-domain to avoid detector latching due to the strong laser excitation pulse \cite{johnston2024cavity}.

\begin{figure*}[!t]
    \centering
    \includegraphics[scale=1]{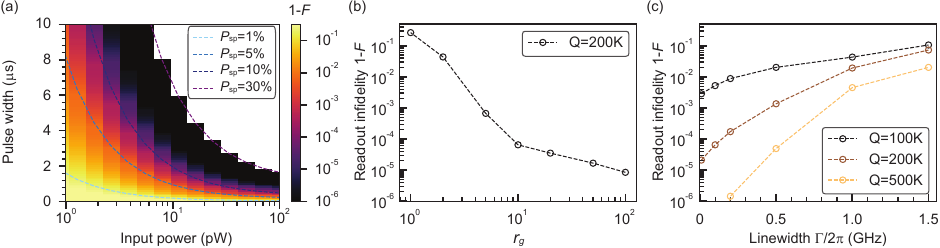}
     \caption{Spin-dependent cavity reflection readout.
     (a) Spin readout infidelity calculations under different laser input powers and pulse widths with optimized $\Delta_a$ and $\Delta_c$ maximizing the reflection contrast. Lower laser powers necessitate longer pulses for reaching the maximal fidelity. The dashed lines label the position where $P_\text{sp}$ population is spin pumped. We ignore conditions that have $P_\text{sp} > 30\%$. System parameters used are listed in Table~\ref{tab:standardParameters}.
     (b) Extracted minimum readout infidelity in panel (a) with different $r_g$ while keeping other system parameters the same. Saturation behavior happens for $r_g \geq 5$.
     (c) The maximum readout fidelity under different system parameters ($Q$, $\Gamma$) with $t_\text{pulse} \leq$ 10 $\mu$s and $1\leq P_\text{in} \leq 100$~pW. 
     }
    \label{fig:Fig3_reflectionReadout}
\end{figure*}

The number of collected fluorescence photons ($N_\text{ph}$) can be calculated as,
\vspace{-5.5pt}
\begin{equation}
    N_\text{ph}=\beta_\text{cav}\eta_\text{sys}\sum_{n=0}^{N_\text{cyc}-1} P_{e}(P_g)^n, 
    \label{eq:fluorescentReadoutFigureOfMerit}
\end{equation}
where $P_e$ is the excited-state population at the end of the laser excitation pulse ($t = t_\text{pulse}$) and $P_g$ is the ground-state population at the end of the first excite-collect sequence ($t=t_\text{seq}$), assuming a perfect initialization process. Note that $P_g$ and $P_e$ correspond to states within a spin-conserving transition defined by the initialized spin state. $\beta_\text{cav}=(\Gamma_\text{cav}-\Gamma_0)/\Gamma_\text{cav}$ is the portion of \textit{T}-center emission coupled to the cavity mode \cite{johnston2024cavity}, with $\Gamma_\text{cav}$ and $\Gamma_0$ being the \textit{T}-center fluorescence decay rate with and without cavity coupling, respectively.  
$\eta_\text{sys} = \eta_\text{cav}\eta_\text{det}= 13.8\%$ is the total system efficiency \cite{johnston2024cavity}, where $\eta_\text{det}=27.5\%$ contains efficiencies for grating coupler, fiber network, and single photon detection.
Note that we target at $\eta_\text{cav} =50\%$ for maximal outcoupling of \textit{T}-center fluorescence \cite{goto2019figure}. Other relevant system parameters are listed in Table~\ref{tab:commonParametersFluorescentReadout}.  
\begin{table}[h]%
    \caption{\label{tab:commonParametersFluorescentReadout}%
    Global parameters considered in the readout fidelity calculation. We use the lower bound of \textit{T}-center quantum efficiency \cite{johnston2024cavity}. $g_e=2$ and $g_h=3.45$ are the ground- and excited-state $g$ factors, respectively \cite{bergeron2020silicon}; $\mu_B$ is the Bohr magneton and $B=200$ mT along silicon [110] is the external magnetic field. The  atom-cavity coupling $g= g_\text{sim}\sqrt{\eta_{QE}}$ assuming ideal \textit{T}-center positioning and dipole alignment with the cavity field. The ideal coupling $g_\text{sim}$ can be achieved via focused-ion-beam implantation to locate single \textit{T} centers at the cavity center, together with photonic cavity fabrication along different silicon crystallographic orientations for better dipole alignment with the cavity polarization.
    }
    \begin{ruledtabular}
        \begin{tabular}{ccc}
            \textrm{Symbol} &
            \textrm{Value} &
            \textrm{Description}\\
            \colrule
            $|g_e-g_h|\mu_BB$     &   4 GHz &      \text{Splitting between transition \textit{A} and \textit{B}} \\
            $g_\text{sim} /2\pi$ & $376$ MHz & \text{Simulated coupling strength} \\
            $\eta_\text{QE}$ & $0.234 $ & \text{Quantum efficiency} \\
        \end{tabular}
    \end{ruledtabular}
\end{table}

Due to the finite cyclicity, the initialized ground state population will eventually be optically pumped away with many excitation cycles, leading to the decrease of fluorescence counts per excitation pulse [Fig.~\ref{fig:Fig2_fluorescenceReadout}(b), inset]. The decay is described by the term $(P_g)^n \approx \exp[-(1-P_g)n]$, considering $P_g \approx 1$. We thus choose the sequence repetition cycle $N_\text{cyc} = (1-P_g)^{-1}$. We define cyclicity of the \textit{T} center as $\Lambda = 1+\Gamma_A/\Gamma_D= 1+\Gamma_B/\Gamma_C$ (Appendix \ref{sec:leve1_Appendix_cyclicity}). The $N_\text{cyc}$ can be related to cyclicity $\Lambda$ via $\Lambda = N_\text{cyc}P_\text{ex}$, where $P_\text{ex}$ is the probability to excite the \textit{T} center in each pulse \cite{raha2020optical}. To simplify the fidelity calculation, we assume that within $N_\text{cyc}$ excitations, the spin remains at its initialized state. Under this assumption, the collected photons during each readout are described by a binomial distribution, which can be further approximated with a Poissonian distribution owing to the low total detection efficiency.

Considering the large splitting between the two spin-conserving transitions (4 GHz) against the cavity linewidth used in the calculation and the long spin $T_1$ lifetime for \textit{T} centers \cite{bergeron2020silicon}, we ignore the minor fluorescence contribution resulted from the optically pumped away population. We choose a relatively low optical excitation power $P_\text{in} = 100$ pW, corresponding to $\sim$ 0.1$P_\text{sat}$ with $P_\text{sat}$ being the saturation pump power, to minimize laser-induced excited state spin mixing \cite{bowness2025laser} and spectral diffusion \cite{bowness2025laser, zhang2025laser}. The mean cavity photon number for $P_\text{in}=$ 100 pW and $Q=200 000$ is estimated by $N_\text{cav} = 4 (\eta_{\text{cav}}/\kappa) (P_{\text{in}}/\hbar\omega_L) = 0.19$, assuming an empty cavity.

Given the cavity and laser frequency location, the measurement of spin-dependent fluorescence yields two Poissonian distributions [$P(N_\text{ph},k)$] with different mean photon number ($N_\text{ph}$) depending on the initialized spin state [Fig.~\ref{fig:Fig2_fluorescenceReadout}(b)]. The spin readout fidelity is given by
\begin{multline} 
    F=\frac{1}{2}\Bigr[P(N_\text{ph}^{\ket{\upg}},k<M) + 0.5P(N_\text{ph}^{\ket{\upg}};k=M) \\
    + P(N_\text{ph}^{\ket{\downg}},k>M) 
    +0.5P(N_\text{ph}^{\ket{\downg}},k=M)\Bigr],
    \label{eq:Fidelity}        
\end{multline}
where $N_\text{ph}^{\ket{\upg}}$ and $N_\text{ph}^{\ket{\downg}}$ are the mean detected photon numbers when the spin state is initialized to $\ket{\upg}$ and $\ket{\downg}$, respectively; $M$ is the photon number threshold that maximize the readout fidelity. Figures \ref{fig:Fig2_fluorescenceReadout}(c) and \ref{fig:Fig2_fluorescenceReadout}(d) summarize spin readout fidelity via resonant fluorescence, depending on system parameters of $r_g$, $\Gamma$, $Q$, and $t_\text{pulse}$. High $r_g$ facilitates more cyclic optical transitions, leading to better fluorescence contrast and thus higher spin readout fidelity. The cyclicity is further boosted by narrower optical cavity (i.e. higher $Q$) and \textit{T}-center optical transition linewidths, which enable larger Purcell enhancement contrast between the resonant transition \textit{A} and other detuned transitions. Longer optical excitation pulse width significantly deteriorates the fidelity, which results from spin pumping via the spin-flipping fluorescence decay ($\Gamma_0/2$) during the optical excitation process.
A short optical excitation ($t_\text{pulse}\leq \sim30$ ns) is required to minimize such an effect. 

With realistic near-term system parameters shown in Table~\ref{tab:standardParameters}
and laser excitation pulses with $t_\text{pulse} = 10$ ns and $P_\text{in} = 100$ pW, spin readout fidelity $F=99.96\%$ can be achieved with a readout time of 179 $\mu$s. By choosing $t_\text{wait} = 7\tau_\text{cav}^\text{on}$, where $\tau_\text{cav}^\text{on}$ is the cavity-enhanced fluorescence lifetime for the transition that is resonant with the cavity, a much faster readout time of 8.7 $\mu$s can be realized with $F$ = 99.97\%. We note that collecting resonant fluorescence with a short decay lifetime can be technically challenging due to the filtering of the laser excitation.

\section{\label{sec:level2_RelectionReadout}Spin readout via cavity Reflection}

Next, we turn to the protocol based on spin-dependent cavity reflection for spin readout. The single-sided cavity output field is described by the input-output relation \cite{gardiner1985input}
\begin{equation}
    \boldOpLe{a}_\text{out} = \boldOpLe{a}_\text{in} + \sqrt{\kappa_\text{wg}} \boldOpLe{a}, 
    \label{eq:inputOutputRelation}
\end{equation}
where $\boldOpLe{a}_\text{in}$ and $\boldOpLe{a}_\text{out}$ are the cavity-coupled input and output laser fields, respectively. The input laser field is given by $\boldOpLe{a}_\text{in} = i\sqrt{\epsilon}$. We compute the cavity reflectivity numerically by solving the master equation for given detunings and \textit{T}-center spin states using
\begin{equation}
    \label{eq:cavityReflection}
    R=\frac{\braket{\boldOpLe{a}^{\dagger}_\text{out}\boldOpLe{a}_\text{out}}}{\braket{\boldOpLe{a}^{\dagger}_\text{in}\boldOpLe{a}_\text{in}}}.
\end{equation}
We note that, due to the \textit{T}-center dephasing, the reflectivity $R\neq|r|^2=|\braket{\boldOpLe{a}_\text{out}}/\braket{\boldOpLe{a}_\text{in}}|^2$ \cite{julsgaard2012reflectivity}. A global optimization \cite{endres2018simplicial} is implemented to locate cavity ($\Delta_c$) and atomic ($\Delta_a$) detunings under a weak excitation condition \cite{waks2006dipole} to maximize the contrast of the spin-dependent cavity reflection (see Appendix~\ref{sec:leve1_Appendix_cavityEfficiency}). The cavity-reflected photon number ($N_\text{ph}$) upon a laser pulse ($t_\text{pulse}$) radiation can be calculated by integrating the photon output flux of the cavity over the time duration of the laser pulse
\begin{equation}
    \begin{split}
    N_\text{ph} 
    =\eta_\text{det}\int_0^{t_\text{pulse}} \text{Tr}\Big[\rho(t)  \boldOpLe{a}^{\dagger}_\text{out}\boldOpLe{a}_\text{out} \Big]dt,
    \label{eq:reflectionCounts}
    \end{split}
\end{equation}
where $\rho(t)$ is the density matrix of the system [Eq.~(\ref{eq:lindbladian})]. Different \textit{T}-center spin states lead to a contrast of average reflected photon numbers, the distributions of which are utilized to infer the spin readout fidelity using Eq.~(\ref{eq:Fidelity}). Dark counts (3 Hz) has been included in the calculation, but they have a negligible contribution due to the small collection time.

We first investigate the spin readout performance with the realistic experimental parameters listed in the Table~\ref{tab:standardParameters}, which leads to an atomic cooperativity $C = 4g^2/(\kappa \Gamma)=1.2$. 
We keep $\eta_\text{cav} = 50\%$ the same with that used in the fluorescence readout, which turns out to be the preferred value to achieve the highest readout fidelity (Appendix~\ref{sec:leve1_Appendix_cavityEfficiency}). 
Figure~\ref{fig:Fig3_reflectionReadout}(a) demonstrates the readout infidelity with a square laser pulse width of $t_\text{pulse} \leq$ 10 $\mu$s and laser power $1\leq P_\text{in} \leq 100$ pW. The minimum readout infidelity $1-F = 6.5\times 10^{-5}$ can be achieved with $t_\text{pulse} = 9.2$ $\mu$s and $P_\text{in} = 1.83$ pW with spin pumping $P_\text{sp} \leq 10\%$.
We note that longer pulses and higher powers will cause spin pumping, which can deteriorate the spin readout fidelity performance. 
Lower laser powers necessitate longer pulse width to reach the maximal fidelity (Appendix ~\ref{sec:level1_Appendix_longPulse}). The tradeoff between fidelity and readout duration is needed for different applications. We select $P_\text{sp} \leq 10\%$ as the bound during readout fidelity calculations. We choose a low spin pumping probability to promote quantum nondemolishing spin readout \cite{raha2020optical}.

The cavity reflection-based readout has a less demanding requirement on the ratio $r_g$ [Fig.~\ref{fig:Fig3_reflectionReadout}(b)] compared to the resonance fluorescence-based readout due to the lower power operation and less involvement of the optical excited states. The reflection readout depends strongly on the cooperativity, where higher readout fidelity can be achieved with larger cavity quality factors and narrower \textit{T}-center optical linewidths [Fig.~\ref{fig:Fig3_reflectionReadout}(c)]. With $C > \sim 0.2$, a readout fidelity $>$ 98\% can be obtained.  
Compared to the fluorescence-based readout, the reflection-based readout holds a few technical advantages by mitigating the requirements of generating short laser pulses and fast time-domain filtering of the laser excitations.

\begin{table}[!h]%
    \caption{\label{tab:standardParameters}%
    Near-term targeted system parameters. For 1D photonic crystal nanobeam cavity, the $Q \sim 8\times10^5$ has been demonstrated \cite {xie2021efficient}. Our champion devices have reached cavity $Q \sim 2\times10^5$. For optical linewidth,\textit{T}-center ensemble showed an inhomogeneous linewidth of 33 MHz in enriched bulk silicon \cite{bergeron2020silicon}. For device-coupled single \textit{T} centers, we observe a narrowest optical linewidth of about 0.5 GHz. These lead to an expectation of 0.1 GHz optical linewidth in the near term with proper material and device engineering (see Sec.~\ref{sec:level1_Conclusion}). The $r_g$ is estimated with spin-selected Purcell enhancement. The highest Purcell factor $F = 44$ we observe on a single \textit{T} center (unpublished) leads to $r_g \sim \sqrt{F} = 7$, assuming the intrinsic branching ratio is 1 \cite{higginbottom2022optical}. We expect $r_g = 10$ can be accessible in the near-term with further improvement of the cavity $Q$ and atom-cavity coupling $g$. 
    }
    \begin{ruledtabular}
        \begin{tabular}{ccc}
            \textrm{Symbol} &
            \textrm{Value} &
            \textrm{Description}\\
            \colrule
            $Q$     &   $2\times10^5$ &      \text{Cavity $Q$ factor} \\ 
            $\Gamma/2\pi$   &   $0.1$ GHz  &   \text{Optical linewidth} \\
            $r_g$ & $10$ & \text{Coupling strength ratio} \\
        \end{tabular}
    \end{ruledtabular}
\end{table}

\section{\label{sec:level3_SpectralDiffusion}Optical dephasing and Spectral diffusion}

In this section, we investigate the influence of the spectral diffusion on the two spin readout protocols. Even though \textit{T}-center emission stays stable in the dark up to $\sim$ ms, they have been shown to suffer from laser-induced spectral diffusion \cite{zhang2025laser, bowness2025laser}. Based on photon correlation measurements, a single \textit{T}-center spectral diffusion timescale has been measured on the order of tens of $\mu$s \cite{bowness2025laser, johnston2024cavity}. We include \textit{T}-center optical transition broadening due to the pure dephasing for both excited states $\ket{\downe}$ and $\ket{\upe}$ via the jump operators $\sqrt{\Gamma_\text{d}/2}\boldOpSy{\sigma}_{z1}$ and $\sqrt{\Gamma_\text{d}/2}\boldOpSy{\sigma}_{z2}$, respectively, in the master equation beyond the lifetime limited linewidth ($\Gamma_0$), revealing dephasing-broadened linewidth $\Gamma = \Gamma_0+2\Gamma_\text{d}$. Considering the spectral diffusion timescale is comparable to or longer than the repetition time of a readout measurement, we simplify the diffusion process during spin readout by assuming $\omega_a = \omega_a^0+\delta\omega$, with $\omega_a^0$ being the average of the bare atomic transition and $\delta\omega$ being a random variable which may be different for each measurement shot. The distribution of the detected photon counts can be evaluated~as,
\begin{equation}
    \text{Dist}(k) = \int P[\mu(\delta\omega),k] G(\delta\omega) d\delta\omega,
    \label{eq:distributionWithdiffusion}
\end{equation}
where $P[\mu(\delta\omega),k]$ is the Poissonian distribution with a mean photon count $\mu(\delta\omega)$, depending on the spectral diffusion value $\delta\omega$ at a specific measurement shot. The spectral diffusion is modeled as a Gaussian distribution
\begin{equation}
    G(\delta\omega) = \frac{1}{\Gamma_\text{sd}}\sqrt{\frac{\text{ln}2}{\pi}}\exp\left[-\text{ln}2 \left( \frac{\delta\omega}{\Gamma_\text{sd}} \right)^2 \right], 
    \label{eq:gaussiandistribution}
\end{equation}
where 2$\Gamma_\text{sd}$ is the full width at half maximum (FWHM) of the broadening due to the spectral diffusion. 

\begin{figure}[t]
    \centering
    \includegraphics[scale=1]{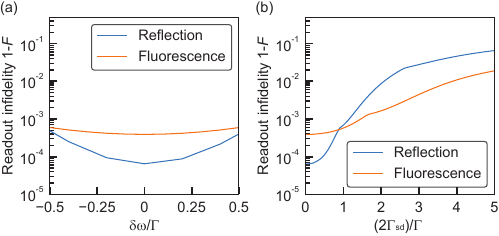}
     \caption{Effect of spectral diffusion on spin readout. 
     (a) Readout infidelity at different optical transition detunings ($\delta\omega$) in two protocols. For each spectral detuning $\delta\omega$, all other system and readout parameters are kept the same as the case when $\delta\omega =0$ in the calculation. 
     (b) Readout infidelity for the two protocols under the spectral diffusion manifested as a random spectral wandering, which is modeled as a Gaussian distribution with a FWHM of 2$\Gamma_\text{sd}$. Gaussian-weighted average of the detected photon distributions at different $\delta\omega$ are used to calculate the spin readout fidelity. For both panels (a) and (b), the results are derived with cQED parameters listed in Table~\ref{tab:standardParameters}. Fluorescence-based readout utilizes $t_\text{pulse} = 10$ ns and $P_\text{in} = 100$ pW while the reflection-based readout uses $t_\text{pulse} = 9.2$ $\mu$s and $P_\text{in} = 1.83$ pW. Both $\delta\omega$ and $2\Gamma_\text{sd}$ are presented in units of the fast dephasing-broadened linewidth $\Gamma/2\pi = 0.1$ GHz.
     }
    \label{fig:Fig4_SpectralDiffusion}
\end{figure}

The spectral diffusion of the \textit{T}-center optical transition causes the readout infidelity to increase in both protocols [Fig.~\ref{fig:Fig4_SpectralDiffusion}(a)]. In the fluorescence-based readout, detuning between the cavity and the spin-conserving transition lowers Purcell enhancement and subsequent fluorescence counts. In the reflection-based readout, deviation of $\Delta_{ac} = \Delta_a-\Delta_c$ from its optimal value decreases the reflection contrast, causing the increase of readout infidelity. 
Under the spectral diffusion, the detected photon count distributions for different initialized spin states are obtained by the Gaussian-weighted
averages of the readout trials with different $\delta\omega$ [Eqs.~(\ref{eq:distributionWithdiffusion}) and ~(\ref{eq:gaussiandistribution})]. The same thresholding method [Eq.~(\ref{eq:Fidelity})] is applied to infer the readout fidelity [Fig.~\ref{fig:Fig4_SpectralDiffusion}(b)]. The fluorescence-based readout is more resilient to the spectral diffusion thanks to the large Purcell enhancement contrast between the near-resonant and off-resonant optical transitions.
The reflection-based readout suffers more from the spectral diffusion due to the fast decrease of the cavity reflection contrast beyond the\textit{T}-center cavity-enhanced optical linewidth (Appendix~\ref{sec:leve1_Appendix_cavityEfficiency}).

\section{\label{sec:level1_Conclusion}DISCUSSION AND CONCLUSION}

Now we discuss the limiting factors and pathways to reach the theoretical readout performance in experiments. The fluorescence-based readout requires high transition cyclicity, which can be enhanced by optimizing alignment between the external magnetic field and the cavity polarization \cite{raha2020optical}. To collect the fast resonant fluorescence signal ($<$ $\sim$10 ns) and filter out the laser excitations, spatial filtering can be utilized to improve the existing time-domain filtering method via gating the detector. Both protocols necessitate a narrow \textit{T}-center optical linewidth ($\Gamma$) and a large atom-cavity coupling strength ($g$), as well as a low-loss optical cavity ($\kappa$). For the optical linewidth, electrical field control may be used to minimize spectral diffusion of cavity-coupled \textit{T} centers via depletion of the charge noises. The same technique was demonstrated for divacancies in SiC, achieving linewidth narrowing by a factor of 30  \cite{anderson2019electrical}. The lower power operation ($\sim$ pW) in the cavity reflection-based readout can benefit the optical linewidth by minimizing the environmental charge reconfiguration due to the excitation laser \cite{zhang2025laser, bowness2025laser}. To promote higher coupling constant $g$, we note that focused-ion-beam-based \cite{schroder2017scalable} and masked \cite{toyli2010chip} ion implantation can be leveraged to increase the yield of \textit{T}-center generation at the cavity center.

In summary, we proposed and investigated two protocols for single-shot spin readout of cavity-coupled \textit{T}-center electronic spins via resonant fluorescence and spin-modulated cavity reflection. Both protocols can enable fast readout ($\sim$ 10 $\mu$s) and reach fidelity $F>99\%$ with realistic system parameters, which put them among the state-of-the-art spin readout performance achieved in the solid-state defects \cite{bourassa2020entanglement, bhaskar2020experimental, ourari2023indistinguishable}. The high fidelity single-shot spin readout is an enabling step towards a broad range of \textit{T}-center-spin-based quantum information applications \cite{afzal2024distributed}. The proposed computation framework can be utilized for exploring other cQED-enabled quantum optical control for \textit{T} centers such as cavity-mediated spin-spin interactions \cite{evans2018photon} and cavity-enhanced Raman emission \cite{sun2018cavity}. 

\begin{center}{
\textbf{\small{DATA AVAILABILITY}}
}
\end{center}

The data that support the findings of this article are openly available \cite{own2026Dataset}.

\begin{figure*}[t]
    \centering
    \includegraphics[scale=1]{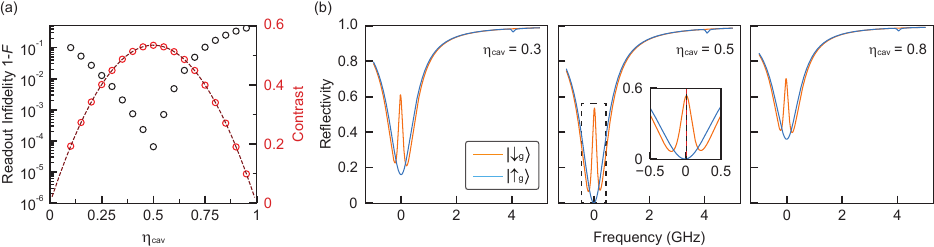}
     \caption{Contrast and fidelity in reflection-based readout with different cavity efficiency $\eta_\text{cav}$.
     (a) QuTip-calculated infidelity (black dots) and optimal spin-dependent reflection contrast (red dots) with system parameters listed in Table.~\ref{tab:standardParameters} and $t_\text{pulse} \leq 10\,\mu$s, $1\leq P_\text{in} \leq 100$~pW. Fidelity and contrast decrease as the cavity deviates from critical coupling. The red dashed line plots the analytical contrast [Eq.~(\ref{eq:analyticalContrast})].
     (b) Analytical reflection curves for varying $\eta_\text{cav}$ using QuTiP optimized $\Delta_a$ and $\Delta_c$. Blue (orange) curves denote initialized spin state to be $\ket{\upg}$ ($\ket{\downg}$). For $\eta_\text{cav} = 0.5$, black and red lines in the inset mark the QuTiP-optimized locations of the optical transition $A$ and the maximal contrast point, respectively.
     }
    \label{fig:Fig5_CavityEfficiency}
\end{figure*}

\begin{acknowledgments}
We gratefully acknowledge Qiyang Huang, Adam Johnston, Shuo Sun, Yizhi Zhu and Geoffroy Hautier for helpful discussions. Support for this research was provided by the National Science Foundation CAREER Award (No. 2238298) and Electronics, Photonics and Magnetic Devices (EPMD) Program (No. 2527905), as well as the Robert A. Welch Foundation (Grant No. C-2134).
\end{acknowledgments}

\appendix

\newpage
\section{TRANSITION CYCLICITY}\label{sec:leve1_Appendix_cyclicity}

We define the optical transition cyclicity for the cavity-coupled \textit{T} center as follows:

\begin{equation}
    \label{eq:cyclicity}
    \Lambda=1+\frac{\Gamma_B}{\Gamma_C}=1 + \frac{\Gamma_A}{\Gamma_D}=1 + \frac{\Gamma_0/2+\frac{4g_{||}^2/\tilde{\kappa}}{1+(2\Delta_{A}/\tilde{\kappa})^2}}{\Gamma_0/2+\frac{4g_{\perp}^2/\tilde{\kappa}}{1+(2\Delta_{D}/\tilde{\kappa})^2}},
\end{equation}
where the $\tilde\kappa=\kappa+2\Gamma_\text{d}$, $\Delta_A$ and $\Delta_D$ are the cavity detunings for the transition $A$ and $D$, respectively. The cavity coupling enhances the cyclicity due to the selective Purcell enhancement of spin-conserving optical decay pathways ($g_{||}^2 \gg g_{\perp}^2$). A small additional enhancement arises from detuning of the spin-non-conserving transitions from the cavity (e.g., $\Delta_A=0, \Delta_D>0$). 

With system parameters of $Q=1\times10^5$ and $\Gamma = 2\pi\times 1$ GHz in the fluorescence readout [$\Delta_A=0, \Delta_D=5.6$ GHz, as shown in Fig.~\ref{fig:Fig2_fluorescenceReadout}(b)], we calculate $\Lambda = 348 $ via Eq.~(\ref{eq:cyclicity}). Under the same parameter set we have \textit{T}-center excitation probability $P_\text{ex}=0.18$ after each excitation pulse, and $N_\text{cyc} = 811$ [inset of Fig.~\ref{fig:Fig2_fluorescenceReadout}(b)]. We note that $\Lambda < N_\text{cyc}P_\text{ex}$, which is mainly due to the spin pumping effect during the excitation pulse.

\section{REFLECTION CONTRAST AT DIFFERENT CAVITY EFFICIENCY\label{sec:leve1_Appendix_cavityEfficiency}}

Cavity efficiency $\eta_\text{cav} = \kappa_\text{wg}/\kappa$ quantifies cavity decay rate through the waveguide mode over the total cavity decay, which can be tuned during nanofabrication. $\eta_\text{cav}$ determines how efficiently the input laser field is coupled to the cavity mode and subsequently interacts with the \textit{T} center. In this appendix, we look into the influence of the $\eta_\text{cav}$ on the performance of the reflection-based spin readout with system parameters listed in Table ~\ref{tab:standardParameters}.
First, we analyze the maximal fidelity based on reflection contrast [Fig.~\ref{fig:Fig5_CavityEfficiency}(a)] via a global optimization process \cite{endres2018simplicial} of $\Delta_a$ and $\Delta_c$. The maximal contrast is achieved when the cavity and the laser are roughly aligned (within 0.1$\Gamma$) with one of the spin-conserving transitions (e.g., $A$) with $\Delta_a \approx \Delta_g - \Delta_e$ and $\Delta_c \approx 0$.

To explain the trend of contrast and fidelity decrease when $\eta_\text{cav}$ deviates from 0.5, we turn to an analytical model for cavity reflection under the weak excitation condition for a two-level system \cite{waks2006dipole, julsgaard2012reflectivity}. We note that with $r_g =10$, the term $g^2_\perp \approx 0.01 g^2$, which effectively decouples the spin-non-conserving transitions and simplifies the model into two-level systems at different transition frequencies. The reflection can be calculated as
\begin{equation}
    r=\frac{\braket{\boldOpLe{a}_\text{out}}}{\braket{\boldOpLe{a}_\text{in}}} \approx 1-\frac{\kappa\eta_\text{cav}}{\kappa/2 + i\Delta_c + \dfrac{g^2}{i\Delta_a+\Gamma/2}},
    \label{eq:analyiticalReflectionCoeff}
\end{equation}
\begin{equation}
    R=|r|^2+ \frac{\kappa^2 \eta^2_\text{cav}}{\kappa^2/4+\Delta_c^2} 
    \frac{1}{\left| 1+\nu\right|^2}\left(\frac{ \braket{\boldOpLe{a^{\dagger}a}} }{|\!\braket{\boldOpLe{a}}\!|^2}-1\right),
    \label{eq:analyiticalReflection}
\end{equation}
where
\begin{equation}
    \nu=\dfrac{g^2}{(\kappa/2+i\Delta_c)(\Gamma/2+i\Delta_a)},
    \label{eq:coeffnu}
\end{equation}
\begin{equation}
    \frac{ \braket{\boldOpLe{a^{\dagger}a}} }{|\!\braket{\boldOpLe{a}}\!|^2}=1+\frac{2g^4(\kappa+\Gamma)\Gamma_\text{d}}{D}\frac{1}{\left(\Gamma/2\right)^2+\Delta_a^2},
    \label{eq:coherentRatio}
\end{equation}
\begin{equation}
    D=\kappa\Gamma_0\Big[(\kappa+\Gamma)^2/4+\Delta_{ac}^2\Big]
    + g^2(\kappa+\Gamma)(\kappa+\Gamma_0).
    \label{eq:coeffD}
\end{equation}

\noindent In the absence of phase noise ($\Gamma_\text{d}=0$), the cavity field becomes coherent with $\braket{\boldOpLe{a^{\dagger}a}} = |\!\braket{\boldOpLe{a}}\!|^2$. This simplifies the cavity reflectivity $R=|r|^2$. In the \textit{T}-center cQED system with nonzero $\Gamma_\text{d}$, the \textit{T}-center dephasing can introduce incoherent cavity field, leading to $R>|r|^2$. Figure~\ref{fig:Fig5_CavityEfficiency}(b) shows the analytically calculated spin-dependent cavity reflection under the two-level system approximation using Eq.~(\ref{eq:analyiticalReflection}) to~(\ref{eq:coeffD}).

Using the QuTiP optimized $\Delta_a$ and $\Delta_c$, we can obtain an analytical expression for the reflection contrast as
\begin{equation}
    \text{Contrast}= 2.141 \eta_\text{cav} - 2.144 \eta^2_\text{cav}.
    \label{eq:analyticalContrast}
\end{equation}
The analytical parabolic functional shape and the maximal contrast of 0.535 (happens at $\eta_\text{cav} = 0.499$) match very well with the numerical results [Fig.~\ref{fig:Fig5_CavityEfficiency}(a)]. The larger contrast helps improve the readout fidelity under the same laser pulse width, power, and spin-pumping limits. We note that $\eta_\text{cav} = 0.5$ can further boost the readout fidelity due to the fact that one spin state gives zero reflection. The slight asymmetry in the infidelity profile is caused by the finite optical pumping effect.

In quantum information applications, critically coupled cavity ($\eta_\text{cav} = 0.5$) can improve the spin-photon entanglement fidelity \cite{bhaskar2020experimental, nguyen2019integrated} while over coupled cavity ($\eta_\text{cav} > 0.5$) can enable efficient spin-dependent phase flip for spin readout \cite{stas2022robust}.

\section{LONGER LASER PULSE IN REFLECTION-BASED READOUT\label{sec:level1_Appendix_longPulse}}   

\begin{figure}[h!]
    \centering
    \includegraphics[scale=1]{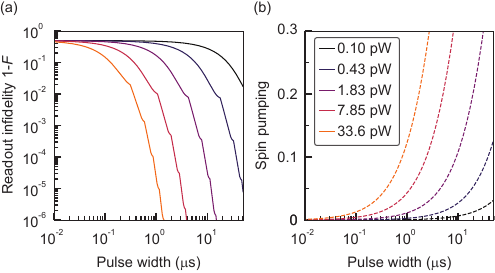}
     \caption{
     \textbf{Reflection-based readout with longer laser pulses.}
     \textbf{(a)} Calculated readout infidelity for reflection-based readout with different laser powers and pulse widths. 
     Lower laser power requires a longer laser pulse to accumulate sufficient photon counts for reaching the optimal fidelity. 
     \textbf{(b)} Spin pumping during the readout sequence. Larger optical power and longer pulse width inevitably lead to optical pumping of the spin state. Cavity efficiency is kept at $\eta_\text{cav} = 0.5$ and all other system parameters are specified in Table.~\ref{tab:standardParameters} for both plots. 
     }
    \label{fig:Fig6_LongPulse}
\end{figure}

In the reflection-based readout (Sec.~\ref{sec:level2_RelectionReadout}), given the system parameters, the input laser power and pulse width determine the readout fidelity [Fig.~\ref{fig:Fig3_reflectionReadout}(a)], where the pulse width is chosen to be bounded below 10 $\mu$s to enable fast single-shot readout. Here, we investigate how longer laser pulse width can affect the readout performance [Fig.\ref{fig:Fig6_LongPulse}(a)]. For a specific laser power, longer laser pulse helps accumulate reflected photons, which promotes the readout fidelity given the reflection contrast.
However, due to nonnegligible excitations, a longer probing laser can induce spin flips, as shown in [Fig. \ref{fig:Fig6_LongPulse}(b)]. In a critically coupled cavity, the off-resonant spin state has a near-zero reflectivity, which makes the reflection protocol robust under optical pumping. Based on different applications, the non-demolishing spin readout may be required to maintain the spin information. A tradeoff needs to be made between readout fidelity, readout time, and spin pumping for different applications.

\newpage
\bibliography{TReadout_references}

@article{jones1973temperature,
  title={Temperature, stress, and annealing effects on the luminescence from electron-irradiated silicon},
  author={Jones, Colin E and Johnson, Eric S and Compton, W Dale and Noonan, JR and Streetman, BG},
  journal={J. Appl. Phys.},
  volume={44},
  number={12},
  pages={5402--5410},
  year={1973},
  publisher={American Institute of Physics}
}

@article{gardiner1985input,
  title={Input and output in damped quantum systems: Quantum stochastic differential equations and the master equation},
  author={Gardiner, Crispin W and Collett, Matthew J},
  journal={Phys. Rev. A},
  volume={31},
  number={6},
  pages={3761},
  year={1985},
  publisher={APS}
}

@article{waks2006dipole,
  title={Dipole induced transparency in drop-filter cavity-waveguide systems},
  author={Waks, Edo and Vuckovic, Jelena},
  journal={Phys. Rev. Lett.},
  volume={96},
  number={15},
  pages={153601},
  year={2006},
  publisher={APS}
}

@article{togan2010quantum,
  title={Quantum entanglement between an optical photon and a solid-state spin qubit},
  author={Togan, Emre and Chu, Yiwen and Trifonov, Alexei S and Jiang, Liang and Maze, Jeronimo and Childress, Lilian and Dutt, MV Gurudev and S{\o}rensen, Anders S{\o}ndberg and Hemmer, Phillip R and Zibrov, Alexander S and others},
  journal={Nature},
  volume={466},
  number={7307},
  pages={730--734},
  year={2010},
  publisher={Nature Publishing Group UK London}
}

@article{vamivakas2010observation,
  title={Observation of spin-dependent quantum jumps via quantum dot resonance fluorescence},
  author={Vamivakas, A Nick and Lu, C-Y and Matthiesen, C and Zhao, Ya and F{\"a}lt, S and Badolato, A and Atat{\"u}re, M},
  journal={Nature},
  volume={467},
  number={7313},
  pages={297--300},
  year={2010},
  publisher={Nature Publishing Group UK London}
}

@article{toyli2010chip,
  title={Chip-scale nanofabrication of single spins and spin arrays in diamond},
  author={Toyli, David M and Weis, Christoph D and Fuchs, Gregory D and Schenkel, Thomas and Awschalom, David D},
  journal={Nano Lett.},
  volume={10},
  number={8},
  pages={3168--3172},
  year={2010},
  publisher={ACS Publications}
}

@article{robledo2011high,
  title={High-fidelity projective read-out of a solid-state spin quantum register},
  author={Robledo, Lucio and Childress, Lilian and Bernien, Hannes and Hensen, Bas and Alkemade, Paul FA and Hanson, Ronald},
  journal={Nature},
  volume={477},
  number={7366},
  pages={574--578},
  year={2011},
  publisher={Nature Publishing Group UK London}
}

@article{gao2012observation,
  title={Observation of entanglement between a quantum dot spin and a single photon},
  author={Gao, WB and Fallahi, Parisa and Togan, Emre and Miguel-S{\'a}nchez, Javier and Imamoglu, Atac},
  journal={Nature},
  volume={491},
  number={7424},
  pages={426--430},
  year={2012},
  publisher={Nature Publishing Group UK London}
}

@article{julsgaard2012reflectivity,
  title={Reflectivity and transmissivity of a cavity coupled to two-level systems: Coherence properties and the influence of phase decay},
  author={Julsgaard, Brian and M{\o}lmer, Klaus},
  journal={Phys. Rev. A},
  volume={85},
  number={1},
  pages={013844},
  year={2012},
  publisher={APS}
}

@article{de2012quantum,
  title={Quantum-dot spin--photon entanglement via frequency downconversion to telecom wavelength},
  author={De Greve, Kristiaan and Yu, Leo and McMahon, Peter L and Pelc, Jason S and Natarajan, Chandra M and Kim, Na Young and Abe, Eisuke and Maier, Sebastian and Schneider, Christian and Kamp, Martin and others},
  journal={Nature},
  volume={491},
  number={7424},
  pages={421--425},
  year={2012},
  publisher={Nature Publishing Group UK London}
}

@article{johansson2012qutip,
  title={QuTiP: An open-source Python framework for the dynamics of open quantum systems},
  author={Johansson, J Robert and Nation, Paul D and Nori, Franco},
  journal={Comput. Phys. Commun.},
  volume={183},
  number={8},
  pages={1760--1772},
  year={2012},
  publisher={Elsevier}
}

@article{bernien2013heralded,
  title={Heralded entanglement between solid-state qubits separated by three metres},
  author={Bernien, Hannes and Hensen, Bas and Pfaff, Wolfgang and Koolstra, Gerwin and Blok, Machiel S and Robledo, Lucio and Taminiau, Tim H and Markham, Matthew and Twitchen, Daniel J and Childress, Lilian and others},
  journal={Nature},
  volume={497},
  number={7447},
  pages={86--90},
  year={2013},
  publisher={Nature Publishing Group UK London}
}

@article{pfaff2014unconditional,
  title={Unconditional quantum teleportation between distant solid-state quantum bits},
  author={Pfaff, Wolfgang and Hensen, Bas J and Bernien, Hannes and van Dam, Suzanne B and Blok, Machiel S and Taminiau, Tim H and Tiggelman, Marijn J and Schouten, Raymond N and Markham, Matthew and Twitchen, Daniel J and others},
  journal={Science},
  volume={345},
  number={6196},
  pages={532--535},
  year={2014},
  publisher={American Association for the Advancement of Science}
}

@article{delteil2014observation,
  title={Observation of quantum jumps of a single quantum dot spin using submicrosecond single-shot optical readout},
  author={Delteil, Aymeric and Gao, Wei-Bo and Fallahi, Parisa and Miguel-Sanchez, Javier and Imamo{\u{g}}lu, Atac},
  journal={Phys. Rev. Lett.},
  volume={112},
  number={11},
  pages={116802},
  year={2014},
  publisher={APS}
}

@article{shields2015efficient,
  title={Efficient readout of a single spin state in diamond via spin-to-charge conversion},
  author={Shields, Brendan J and Unterreithmeier, Quirin P and de Leon, Nathalie P and Park, H and Lukin, Mikhail D},
  journal={Phys. Rev. Lett.},
  volume={114},
  number={13},
  pages={136402},
  year={2015},
  publisher={APS}
}

@article{sukachev2017silicon,
  title={Silicon-vacancy spin qubit in diamond: a quantum memory exceeding 10 ms with single-shot state readout},
  author={Sukachev, Denis D and Sipahigil, Alp and Nguyen, Christian T and Bhaskar, Mihir K and Evans, Ruffin E and Jelezko, Fedor and Lukin, Mikhail D},
  journal={Phys. Rev. Lett.},
  volume={119},
  number={22},
  pages={223602},
  year={2017},
  publisher={APS}
}

@article{schroder2017scalable,
  title={Scalable focused ion beam creation of nearly lifetime-limited single quantum emitters in diamond nanostructures},
  author={Schr{\"o}der, Tim and Trusheim, Matthew E and Walsh, Michael and Li, Luozhou and Zheng, Jiabao and Schukraft, Marco and Sipahigil, Alp and Evans, Ruffin E and Sukachev, Denis D and Nguyen, Christian T and others},
  journal={Nat. Commun.},
  volume={8},
  number={1},
  pages={15376},
  year={2017},
  publisher={Nature Publishing Group UK London}
}

@article{sun2018cavity,
  title={Cavity-enhanced Raman emission from a single color center in a solid},
  author={Sun, Shuo and Zhang, Jingyuan Linda and Fischer, Kevin A and Burek, Michael J and Dory, Constantin and Lagoudakis, Konstantinos G and Tzeng, Yan-Kai and Radulaski, Marina and Kelaita, Yousif and Safavi-Naeini, Amir and others},
  journal={Phys. Rev. Lett.},
  volume={121},
  number={8},
  pages={083601},
  year={2018},
  publisher={APS}
}

@article{chartrand2018highly,
  title={{Highly enriched Si 28 reveals remarkable optical linewidths and fine structure for well-known damage centers}},
  author={Chartrand, C and Bergeron, L and Morse, K. J. and Riemann, H and Abrosimov, N. V. and Becker, P and Pohl, H-J and Simmons, S and Thewalt, M. L. W},
  journal={Phys. Rev. B},
  volume={98},
  number={19},
  pages={195201},
  year={2018},
  publisher={APS}
}

@article{awschalom2018quantum,
  title={Quantum technologies with optically interfaced solid-state spins},
  author={Awschalom, David D and Hanson, Ronald and Wrachtrup, J{\"o}rg and Zhou, Brian B},
  journal={Nat. Photon.},
  volume={12},
  number={9},
  pages={516--527},
  year={2018},
  publisher={Nature Publishing Group UK London}
}

@article{evans2018photon,
  title={Photon-mediated interactions between quantum emitters in a diamond nanocavity},
  author={Evans, Ruffin E and Bhaskar, Mihir K and Sukachev, Denis D and Nguyen, Christian T and Sipahigil, Alp and Burek, Michael J and Machielse, Bartholomeus and Zhang, Grace H and Zibrov, Alexander S and Bielejec, Edward and others},
  journal={Science},
  volume={362},
  number={6415},
  pages={662--665},
  year={2018},
  publisher={American Association for the Advancement of Science}
}

@article{dibos2018atomic,
  title={Atomic source of single photons in the telecom band},
  author={Dibos, A. M. and Raha, Mouktik and Phenicie, C. M. and Thompson, Jeffrey Douglas},
  journal={Phys. Rev. Lett.},
  volume={120},
  number={24},
  pages={243601},
  year={2018},
  publisher={APS}
}

@article{endres2018simplicial,
  title={A simplicial homology algorithm for Lipschitz optimisation},
  author={Endres, Stefan C and Sandrock, Carl and Focke, Walter W},
  journal={J. Glob. Optim.},
  volume={72},
  pages={181--217},
  year={2018},
  publisher={Springer}
}

@article{anderson2019electrical,
  title={Electrical and optical control of single spins integrated in scalable semiconductor devices},
  author={Anderson, Christopher P and Bourassa, Alexandre and Miao, Kevin C and Wolfowicz, Gary and Mintun, Peter J and Crook, Alexander L and Abe, Hiroshi and Ul Hassan, Jawad and Son, Nguyen T and Ohshima, Takeshi and others},
  journal={Science},
  volume={366},
  number={6470},
  pages={1225--1230},
  year={2019},
  publisher={American Association for the Advancement of Science}
}

@article{nguyen2019integrated,
  title={An integrated nanophotonic quantum register based on silicon-vacancy spins in diamond},
  author={Nguyen, CT and Sukachev, D. D. and Bhaskar, M. K. and Machielse, Bartholomeus and Levonian, D. S. and Knall, E. N. and Stroganov, Pavel and Chia, Cleaven and Burek, M. J. and Riedinger, Ralf and others},
  journal={Phys. Rev. B},
  volume={100},
  number={16},
  pages={165428},
  year={2019},
  publisher={APS}
}

@article{goto2019figure,
  title={Figure of merit for single-photon generation based on cavity quantum electrodynamics},
  author={Goto, Hayato and Mizukami, Shota and Tokunaga, Yuuki and Aoki, Takao},
  journal={Phys. Rev. A},
  volume={99},
  number={5},
  pages={053843},
  year={2019},
  publisher={APS}
}

@article{bourassa2020entanglement,
  title={Entanglement and control of single nuclear spins in isotopically engineered silicon carbide},
  author={Bourassa, Alexandre and Anderson, Christopher P and Miao, Kevin C and Onizhuk, Mykyta and Ma, He and Crook, Alexander L and Abe, Hiroshi and Ul-Hassan, Jawad and Ohshima, Takeshi and Son, Nguyen T and others},
  journal={Nat. Mater.},
  volume={19},
  number={12},
  pages={1319--1325},
  year={2020},
  publisher={Nature Publishing Group UK London}
}

@article{bergeron2020silicon,
  title={Silicon-integrated telecommunications photon-spin interface},
  author={Bergeron, L and Chartrand, C and Kurkjian, A. T. K. and Morse, K. J. and Riemann, H and Abrosimov, N. V. and Becker, P and Pohl, H-J and Thewalt, M. L. W. and Simmons, S},
  journal={PRX Quantum},
  volume={1},
  number={2},
  pages={020301},
  year={2020},
  publisher={APS}
}

@article{bhaskar2020experimental,
  title={Experimental demonstration of memory-enhanced quantum communication},
  author={Bhaskar, Mihir K and Riedinger, Ralf and Machielse, Bartholomeus and Levonian, David S and Nguyen, Christian T and Knall, Erik N and Park, Hongkun and Englund, Dirk and Lon{\v{c}}ar, Marko and Sukachev, Denis D and others},
  journal={Nature},
  volume={580},
  number={7801},
  pages={60--64},
  year={2020},
  publisher={Nature Publishing Group UK London}
}

@article{kindem2020control,
  title={Control and single-shot readout of an ion embedded in a nanophotonic cavity},
  author={Kindem, Jonathan M and Ruskuc, Andrei and Bartholomew, John G and Rochman, Jake and Huan, Yan Qi and Faraon, Andrei},
  journal={Nature},
  volume={580},
  number={7802},
  pages={201--204},
  year={2020},
  publisher={Nature Publishing Group UK London}
}

@article{raha2020optical,
  title={Optical quantum nondemolition measurement of a single rare earth ion qubit},
  author={Raha, Mouktik and Chen, Songtao and Phenicie, Christopher M and Ourari, Salim and Dibos, Alan M and Thompson, Jeff D},
  journal={Nat. Commun.},
  volume={11},
  number={1},
  pages={1605},
  year={2020},
  publisher={Nature Publishing Group}
}

@article{xie2021efficient,
  title={{Efficient resonance management in ultrahigh-Q 1D photonic crystal nanocavities fabricated on 300 mm SOI CMOS platform}},
  author={Xie, Weiqiang and Verheyen, Peter and Pantouvaki, Marianna and Van Campenhout, Joris and Van Thourhout, Dries},
  journal={Laser \& Photonics Reviews},
  volume={15},
  number={2},
  pages={2000317},
  year={2021},
  publisher={Wiley Online Library}
}

@article{stas2022robust,
  title={Robust multi-qubit quantum network node with integrated error detection},
  author={Stas, P-J and Huan, Yan Qi and Machielse, Bartholomeus and Knall, Erik N and Suleymanzade, Aziza and Pingault, Benjamin and Sutula, Madison and Ding, Sophie W and Knaut, Can M and Assumpcao, Daniel R and others},
  journal={Science},
  volume={378},
  number={6619},
  pages={557--560},
  year={2022},
  publisher={American Association for the Advancement of Science}
}

@article{higginbottom2022optical,
  title={Optical observation of single spins in silicon},
  author={Higginbottom, Daniel B and Kurkjian, Alexander TK and Chartrand, Camille and Kazemi, Moein and Brunelle, Nicholas A and MacQuarrie, Evan R and Klein, James R and Lee-Hone, Nicholas R and Stacho, Jakub and Ruether, Myles and others},
  journal={Nature},
  volume={607},
  number={7918},
  pages={266--270},
  year={2022},
  publisher={Nature Publishing Group UK London}
}

@article{anderson2022five,
  title={Five-second coherence of a single spin with single-shot readout in silicon carbide},
  author={Anderson, Christopher P and Glen, Elena O and Zeledon, Cyrus and Bourassa, Alexandre and Jin, Yu and Zhu, Yizhi and Vorwerk, Christian and Crook, Alexander L and Abe, Hiroshi and Ul-Hassan, Jawad and others},
  journal={Sci. Adv.},
  volume={8},
  number={5},
  pages={eabm5912},
  year={2022},
  publisher={American Association for the Advancement of Science}
}

@article{ourari2023indistinguishable,
  title={{Indistinguishable telecom band photons from a single Er ion in the solid state}},
  author={Ourari, Salim and Dusanowski, {\L}ukasz and Horvath, Sebastian P and Uysal, Mehmet T and Phenicie, Christopher M and Stevenson, Paul and Raha, Mouktik and Chen, Songtao and Cava, Robert J and de Leon, Nathalie P and others},
  journal={Nature},
  volume={620},
  number={7976},
  pages={977--981},
  year={2023},
  publisher={Nature Publishing Group UK London}
}

@article{lee2023high,
  title={{High-efficiency single photon emission from a silicon T-center in a nanobeam}},
  author={Lee, Chang-Min and Islam, Fariba and Harper, Samuel and Buyukkaya, Mustafa Atabey and Higginbottom, Daniel and Simmons, Stephanie and Waks, Edo},
  journal={ACS Photon.},
  volume={10},
  number={11},
  pages={3844--3849},
  year={2023},
  publisher={ACS Publications}
}

@article{islam2023cavity,
  title={{Cavity-enhanced emission from a silicon T center}},
  author={Islam, Fariba and Lee, Chang-Min and Harper, Samuel and Rahaman, Mohammad Habibur and Zhao, Yuqi and Vij, Neelesh Kumar and Waks, Edo},
  journal={Nano Lett.},
  volume={24},
  number={1},
  pages={319--325},
  year={2023},
  publisher={ACS Publications}
}

@article{knaut2024entanglement,
  title={Entanglement of nanophotonic quantum memory nodes in a telecom network},
  author={Knaut, Can M and Suleymanzade, Aziza and Wei, Y-C and Assumpcao, Daniel R and Stas, P-J and Huan, Yan Qi and Machielse, Bartholomeus and Knall, Erik N and Sutula, Madison and Baranes, Gefen and others},
  journal={Nature},
  volume={629},
  number={8012},
  pages={573--578},
  year={2024},
  publisher={Nature Publishing Group UK London}
}

@article{johnston2024cavity,
  title={Cavity-coupled telecom atomic source in silicon},
  author={Johnston, Adam and Felix-Rendon, Ulises and Wong, Yu-En and Chen, Songtao},
  journal={Nat. Commun.},
  volume={15},
  number={1},
  pages={2350},
  year={2024},
  publisher={Nature Publishing Group UK London}
}

@article{afzal2024distributed,
  title={Distributed Quantum Computing in Silicon},
  author={Afzal, Francis and Akhlaghi, Mohsen and Beale, Stefanie J and Bedroya, Olinka and Bell, Kristin and Bergeron, Laurent and Bonsma-Fisher, Kent and Bychkova, Polina and Chaisson, Zachary ME and Chartrand, Camille and others},
  journal={arXiv preprint arXiv:2406.01704},
  year={2024}
}

@article{rosenthal2024single,
  title={Single-Shot Readout and Weak Measurement of a Tin-Vacancy Qubit in Diamond},
  author={Rosenthal, Eric I and Biswas, Souvik and Scuri, Giovanni and Lee, Hope and Stein, Abigail J and Kleidermacher, Hannah C and Grzesik, Jakob and Rugar, Alison E and Aghaeimeibodi, Shahriar and Riedel, Daniel and others},
  journal={Phys. Rev. X},
  volume={14},
  number={4},
  pages={041008},
  year={2024},
  publisher={APS}
}

@article{song2026entanglement,
  title={Entanglement of a nuclear spin qubit register in silicon photonics},
  author={Song, Hanbin and Zhang, Xueyue and Komza, Lukasz and Fiaschi, Niccolo and Xiong, Yihuang and Zhi, Yiyang and Dhuey, Scott and Schwartzberg, Adam and Schenkel, Thomas and Hautier, Geoffroy and others},
  journal={Nat. Nanotechnol.},
  volume={21},
  number={1},
  pages={53--57},
  year={2026},
  publisher={Nature Publishing Group UK London}
}

@article{bowness2025laser,
  title={{Laser-induced spectral diffusion and excited-state mixing of silicon T centers}},
  author={Bowness, Camille and Meynell, Simon A and Dobinson, Michael and Clear, Chloe and Jooya, Kais and Brunelle, Nicholas and Keshavarz, Mehdi and Boos, Katarina and Gascoine, Melanie and Taherizadegan, Shahrzad and others},
  journal={PRX Quantum},
  volume={6},
  number={3},
  pages={030350},
  year={2025},
  publisher={APS}
}

@article{komza2025multiplexed,
  title={Multiplexed color centers in a silicon photonic cavity array},
  author={Komza, Lukasz and Zhang, Xueyue and Song, Hanbin and Tang, Yu-Lung and Wei, Xin and Sipahigil, Alp},
  journal={Optica},
  volume={12},
  number={9},
  pages={1400--1405},
  year={2025},
  publisher={Optica Publishing Group}
}

@article{uysal2025spin,
  title={{Spin-photon entanglement of a single Er$^{3+}$ ion in the telecom band}},
  author={Uysal, Mehmet T and Dusanowski, {\L}ukasz and Xu, Haitong and Horvath, Sebastian P and Ourari, Salim and Cava, Robert J and de Leon, Nathalie P and Thompson, Jeff D},
  journal={Phys. Rev. X},
  volume={15},
  number={1},
  pages={011071},
  year={2025},
  publisher={APS}
}

@article{zhang2025laser,
  title={{Laser-induced spectral diffusion of T centers in silicon nanophotonic devices}},
  author={Zhang, Xueyue and Fiaschi, Niccolo and Komza, Lukasz and Song, Hanbin and Schenkel, Thomas and Sipahigil, Alp},
  journal={PRX Quantum},
  volume={6},
  number={3},
  pages={030351},
  year={2025},
  publisher={APS}
}

@misc{own2026Dataset,
  author = {Wong, Yu-En and Chen, Songtao},
  title = {Dataset openly available},
  howpublished = "\url{https://doi.org/10.7910/DVN/KCVKYF}",
  year = {2026}, 
}

\end{document}